# "Orthotropic Piezoelectricity in 2D Nanocellulose"

Y. García[1], Yasser B. Ruiz-Blanco[1,2]*, Yovani Marrero-Ponce[2] and C. M. Sotomayor-Torres[1,3]*

[1]Catalan Institute of Nanoscience and Nanotechnology (ICN2), Campus UAB, 08193 Bellaterra, Spain

[2]Unit of Computer-Aided Molecular "Biosilico" Discovery and Bioinformatics Research (CAMD-BIR Unit), Faculty of Chemistry-Pharmacy. Universidad Central "Marta Abreu" de Las Villas, 54830 Santa Clara, Cuba.

[3]Instituciò Catalana de Recerca i Estudis Avançats (ICREA), 08010 Barcelona, Spain

ABSTRACT

The control of electromechanical responses within bonding regions is essential to face frontier challenges in nanotechnologies, such as molecular electronics and biotechnology. Here, we present Iβ-nanocellulose as a potentially new orthotropic 2D piezoelectric crystal. The predicted in-layer piezoelectricity is originated on a *sui-generis* hydrogen bonds pattern. Upon this fact and by using a combination of *ab-initio* and *ad-hoc* models, we introduce a description of electrical profiles along chemical bonds. Such developments lead to obtain a rationale for modelling the extended piezoelectric effect originated within bond scales. The order of magnitude estimated for the 2D Iβ-nanocellulose piezoelectric response, ~pm V$^{-1}$, ranks this material at the level of currently used piezoelectric energy generators and new artificial 2D designs. Such finding would be crucial for developing alternative materials to drive emerging nanotechnologies.

# Introduction

Sustainable nanotechnologies require pushing the limits in understanding fundamental science and engineering while meeting the challenge of being environmentally sound and capable to afford present technological demands[1,2]. Cellulose, the most abundant polymer on Earth, is a promising material for green nanotechnology. Nanocellulose (NC), obtained by high level processing of this raw material, is attracting great attention from industrial researchers[3-5]. Such trend relies on its unique optomechanical features and green nature[3,5,6]. NC is meant to influence a diversity of scenarios such as medical devices, reinforcing agents in plastic composites, pharma coatings, and smart textiles[5,7-9]. Here, by introducing 2D Iβ-NC piezoelectric crystals, we open this spectrum to still unexplored applications based on NC electromechanical responses[3,4,10,11].

The development of energy harvesters has become one of the most claimed needs of today[12,13]. This alternative is based on the discovery of piezoelectric (PZ) materials that would enable disruptive technological solutions that may allow to answer emerging needs on sensing and actuating e.g. portable and delocalized energy generators. Piezoelectricity is a phenomenon in which an electric field is generated inside a material subjected to a mechanical strain or vice versa. Recent developments in synthetic organic polymers[14], natural 2D $MoS_2$[15], engineered 2D flexible graphene[16], show the global tendency to cover an area aimed to control PZ sources at the nanoscale. The advantage of our proposal relies on the relevant properties of NC as being biocompatible, flexible, transparent[3-8] and its potentially revealed *in-layer* PZ, which would be an intrinsic feature of the crystalline structure of Iβ-NC.

Undoubtedly, the already successful story of molecular technologies would greatly benefit from the ability to accurately understand the fundamentals of the observed responses on the macro-scale[14]. Previous studies suggests that the electromechanical activity of NC could be managed by its hydrogen bonds (HB)[17-19]. However, despite the developments of methods to characterize HB there is still a limited characterization in terms of mean-field parameters of the electrical features and the electromechanical response of such interactions[14,20]. Then so, our main contribution, from the fundamental point of view, is the development of a theory to study HB based on the interatomic electrical profiles. Our approach permits the understanding of the microscopic origin of the predicted PZ in the proposed 2D Iβ-NC crystals, and at the same time it supports previous electromechanical measurements in raw materials such as cellulose[21] and wood[22]. The new electromechanical approach described in this report would also be essential for future developments intended to facilitate a deeper understanding of life-essential molecular mechanisms occurring in hydrogen-bonded systems like DNA.

Full version is available on: https://www.nature.com/articles/srep34616

# Results

**Hydrogen bonds pattern supporting 2D crystals based on Iβ nanocellulose**

Cellulose exists in four allomorphic forms[23]. The coexistence ratio for each possible crystal variation depends on the cellulose origin. Cellulose-I (native cellulose) crystal may consists of one or two different crystal phases[24], either Iα or Iβ. Our studies are focused on Iβ-nanocellulose (Iβ-NC), which could show the exclusive advantages of quasi 2D-orthotropic molecular crystals, due to its distinctive property of being built with layers held by weak Van der Waals (VdW) interactions[25], in contrast to the stronger HB network *within* the layers as depicted in Fig. 1a. Notably, such HB are roughly aligned to two orthogonal twofold axes of rotational symmetry, see Fig. 1b[23]. The experimental structure sketched in the lower panel was derived from the analysis of synchrotron generated X-rays and neutron fiber diffraction data[23]. The cluster shaded by the rectangular region constitutes a representative system of the different types of HB periodically repeated along the structure. This arrangement suggests that HB could be crucial in the transmission of atomic scale electrical modifications along large-scale 2D-crystal arrangements.

Patterning of Iβ-NC refers to the comprehension and characterization of an array of HB in the crystal, see red-dashed lines in Fig. 1. In consequence, the electromechanical picture of the extended system may be understood and modified by managing such nanoscaled patterns. The Iβ-NC crystal comprises eight HB per unit cell: four intra-chain, controlling torsion and directionality, and four inter-chains (with the neighbour cell) being responsible of holding the chains together, see Fig. 1b. In what follows, these two situations will be referred to as intra- and inter-chain directions, respectively. Such features constitute the necessary scaffold to find ways of tuning these interactions without disrupting the covalent network, i.e. allowing reversible modifications to the crystal structure[19].

To understand how HB could influence a collective response in Iβ-NC films, we consider a single-layer-cluster composed by three parallel chains, see Fig. 2. The HB pattern shows the nearly orthogonal directionality between the inter-chain and intra-chain HB raising the possibility of selective manipulation of each group to drive orthotropic effects in the extended crystal.

**An approach to quantify the electrical nature of hydrogen bonds**

Two major problems emerge when trying to quantify HB. First, there is no universal definition of such interactions and the present ones are at best incomplete[26,27]. The description of the electrical features of interactions taking place within the bonding region might be of benefit to model breakthrough phenomena like the electromechanical response of diverse hydrogen-bonded systems[14]. Second, present numerical methods suffer from uncertainties when describing HB due to the limited consideration of the electron-electron interactions in model systems. Such dispersive forces are one of the main contributors to HB formation and further stabilization[28].

To overcome these two difficulties and to gain further understanding of electromechanical responses in hydrogen bonded materials, we study the electrical profiles within HB. Then, we introduce an *ad-hoc* single-bond descriptor[29], $t_a$, see Supp. Mat. 1, to obtain a mean-field characterization of chemical bonds attending to electrical features. The used computational approach was developed by some us[30], using the advantages of Green Function Methods[28]. The methodology accounts for nonlocal[31] dispersive forces while going beyond Density Functional Theory (DFT) as a frame to develop *ab-initio* calculations[32,33].

Some theoretical approaches derived from Atoms in Molecules Theory have been used to describe electron density features of HB[34,35]. Such descriptions successfully respond to a broad range of demands in molecular research. However, in order to design truly nanoscale devices based on the response of the HB to external electric fields, a characterization in terms of electrical forces would be more revealing and suitable. Consequently, this report introduces a new description of the electric field (**E**) within the bond region.

Stationary **E** vectors, are analytically computed as the derivative of the potential energy[36]. This characterization facilitates the understanding of general hydrogen-bonded material responses in the presence of electrical perturbations due to a direct relationship with forces at the single-bond level[14,37]. This rationale can be extended to covalent interactions within the bond region.

Fig.´s 3 and 4 represent interatomic **E**-fields in the NC slab. The evolution of the interatomic |**E**| along the bond length, is indicated in the upper panels while the progression of the angle defined by **E** and the bond axis (**Φ**), is represented in the lower panels. The analysis of **Φ** indicates that the electric field vectors undergo 180º rotations. This mechanism, which takes place in all the studied chemical bonds, evidences the impossibility of modelling bonds as dipoles for phenomena taking place within the bonding region, see also Supp. Mat. 3.1. This concern was initially raised by Pauling in his original work[38] in 1939 but it has been superseded by other reduced electrostatic models for HB[39,40]. Here, we model a HB as two interacting-positive charges modulated by the surrounding electron density. It is precisely the polarization of such electron cloud the responsible for the *sui generis* strength of HB related to other electrical interactions.

To capture information concerning the electromechanical response of each of the chemical bonds when subjected to external fields we use the $t_a$-descriptors, see results included on Fig.´s 3d and 4d. Such values obtained for $t_a$ provides a frame for electrostatic characterization of chemical bonds, covalent (CB) and HB. The larger the descriptor values, the greater the extension of the deviation of the electron density from bond symmetry, which indicates weaker bonds. Hence the $t_a$ descriptor can be understood as an inverse measure of electrical stability. The higher values obtained for HB agree with the evidences that is possible to drive, by electrical means, the HB easier than CB. Such interpretation is aligned to the single-bond PZ response analysis reported for 2-methyl-4-nitroaniline based systems[14]. The effective tuning of the HB network in Iβ-NC might lead to an in-plane modification of the layer mostly in the inter-chain direction where HB are the dominant force. Given the strength difference between HB and VdW interactions, it is not surprising that the described PZ so far, in wood[22] and cellulose-based materials[21,41], result in the out-of-plane direction dominated by VdW interactions. However, VdW interactions are not locally confined as HB, therefore the possibility to control HB opens a wide spectrum of specific and atomic-scaled phenomena which may influence roughness, hardness, permeability, phase transitions, optical properties and electromechanical responses in NC crystals.

# Intrinsic piezoelectric effect in 2D nanocellulose

To understand how the electromechanical features associated with HB can influence the material at a larger scale, we address the estimation of PZ in the Iβ-NC crystal. It is expected that the existence of an electronic band gap and the lack of a symmetry centre[12] would give rise to such PZ in NC. Here, we aim to unveil the possibility of in-plane PZ based on the electromechanical deformation of HB pattern.

The mechanism we propose to explain PZ in 2D-NC resembles how the interaction with **E**-fields leads to the polarization of electron clouds localized in the HB and thenceforth to the mechanical deformation of the HB pattern that supports the structure. Initially, we calculate the piezoelectric response of individual HB, see Fig. 5a. Note that even when the HB have opposed orientations, both in the inter-chain and the intra-chain directions, this feature does not lead to a cancellation of the PZ effect because of the existence of a net dipole moment, see Fig. 1b and Supp. Mat. 3 for details. Then, the extension of such results to the crystal is interpreted as a collective effect where each HB actuates as an individual PZ precursor.

The existence of finite non-zero minimum values of the electric field modulus ($|\mathbf{E}|_{threshold} \sim 10$ V nm$^{-1}$) along the bond axis, see upper panels in Fig.´s 3 and 4, is associated with the electrostatic inertia and therefore represent a threshold for an static response. The linear relationships between the intensity of the applied **E** and the resulting bond strains agree with the intrinsic piezoelectric character of HB[14] (Fig. 5a). The interpolation to zero of the linear regression functions shows that switching off the electric field leads to the full recovery of the initial bond length. In all cases, we also observe how the reversed polarity of the external **E** inverts the mechanical response. We then calculate the PZ tensor of the extended crystal, details in Methods section, and report the orthogonal components in the plane defined by the crystallographic axes b and c, see Fig.'s 1 and 5b. The non-symmetric behaviour is justified by the intrinsic asymmetries in the HB pattern occurring in NC, and leads us to predict an intrinsic orthotropic character in the PZ response in 2D Iβ-NC. However, the values reported for the PZ coefficients in Fig. 5b will be severely affected by the mechanical constriction inherent to CB in the chain direction, c-axis in Fig.1b, where could be expected a total suppression of the PZ response. Finally, the reported $d_{22}$ value for Iβ-NC in the inter-chain direction, ~pm V$^{-1}$, ranks this material at the level of frequently used bulk piezoelectric materials such as α-quartz[42] ($d_{11}$=2.1 pm V$^{-1}$) and improves values estimated for novel atomic scaled materials like doped graphene[16] ($d_{31}$=0.55 pm V$^{-1}$).

**Discussion**

We show that Iβ-NC[3,4] can behave as a 2D orthotropic PZ crystal by means of structural changes originated in the response of a HB pattern to the electrical stress. We obtain numerical estimations of the overall magnitude of the effect by using *ad-hoc* and *ab-initio* models supported in a mechanism based on single-bond piezoelectricity. The extensions to the crystal scale was done using accepted models for similar HB-based materials[14].

To summarize, we would like to highlight two aspects regarding the advantages and generalizations introduced by our methodology: (i) the method used to approach piezoelectric coefficients and (ii) the quantum chemistry based scheme used to estimate numerical quantities in the system. Concerning the first point, we present a characterization of **E** within the bond region to facilitate new electromechanical understandings of single bond responses as well as for supporting the definition of constraints in further models used to predict PZ response. Regarding the second point, we have applied a methodology previously developed by some of us[43,44] to obtain numerical estimates of this phenomenon. The application of this method have permitted us the use of static and mean field DFT calculations to approach non-equilibrium quantities as **E**-field response. In addition, we introduce a descriptor that quantifies the electrical features of chemical bonds in a single-effective parameter. The analysis shows that the larger the deviations from a Heaviside-type behavior in the **E**-profiles (i.e. larger $t_a$ values) the higher would be the PZ response. From a fundamental point of view, such description also complements the current characterization of chemical bonds based on geometric, energetic and spectroscopic parameters[26]. Thus, the characterization of bonds through the $t_a$ descriptor could be an initial precursor for systematic quantification of the electrical nature of chemical bonds. In addition, the benefit of obtaining an structural descriptor associated to the interactions in response to an electric field, should favor the design of nanoelectronic devices based on the selective electrical tuning of the most susceptible chemical bonds.

Finally, the novel approach here introduced opens routes for understanding yet unexplored regions of the molecular world at both theoretical[12] and experimental levels[45]. The promise of 2D piezoelectricity in NC, as well as the understanding of the atomic origin of the phenomenon, is central to nanotechnology not only because of 2D-NC expected broad technological applications in the area of nanoelectronics, but also because NC is an excellent HB model platform for study the impact of intramolecular interactions like ubiquitous HB. For instance, the electromechanical approach here developed would be essential for future developments intended to progress on atomic control of large-scale hydrogen bonded materials like life essential double-stranded DNA. Our effective approach for a quantification of the electrical forces in chemical bonds, is expected to impact areas such of molecular biology or genetic engineering where relative strengths associated to individual HB mediates processes like protein folding, DNA replication, DNA transcription and gene editing[46-50].

# Methods

**Numerical modelling**

Calculations were carried out with an improved nonlocal DFT[31] method suited to describe atomic systems[28,32]. The approach is specialized in dealing with systems involving polarisable electron densities, as required for an *ab-initio* description of HB interactions. The approach comprises the solution of a quantum mechanical ensemble in vacuum. Thus, we first define a reduced atomic model of a NC sheet, light green rectangular region in Fig. 1b. Then, we introduce the interactions of the real system by means of effective external potentials as well as mean field DFT potentials to account for interatomic effects. Our improved DFT methodology accounts for dispersion forces in a systematic manner. The nonlocal exchange-correlation potential comprises nonlocal Fock exchange in addition to the local exchange via the PW91 functional. Electronic states are described by 6-31g (d,p) basis sets[51]. To quantify how the numerical methods influence the mean values here reported we introduce a water dimer. This reduced scenario facilities a confident framework to assess our developments as well as for setting a comprehensive scaffold for the design and optimization of our *ab-initio* methods, see also Supp. Mat. 2 for further details.

The monoclinic unit cell model for Iβ-NC comprises two sheets containing the origin and central chains with similar HB networks. The space group of the crystal is P21 and the unit cell parameters are[23] a= 7.784 Å, b= 8.201Å, and c (chain direction and the polymer unique axis) = 10.38 Å, and then γ (the monoclinic angle) = 96.5º.

**Computational details**

The code GAUSSIAN09 is used for numerical DFT calculations[51]. For the definition of finite cluster models and for the selection of basis sets, the following reasoning was followed. First, the atomic structure of the minimal finite cluster model for the ideal crystalline NC is chosen as described above. Further extensions of this reduced cluster obey crystallographic criteria[23] and are subjected to a convergence criteria. We consider that our calculations converged in terms of the cluster sizes when fixing all the other parameters, the variation in the target quantity is less than 10%. Similarly, we address the convergence due to the basis set selection, see Supp. Mat. 2.

**Model applied for the calculation of piezoelectric coefficients**

The linear PZ effect is described as a first-order coupling between the macroscopic $\mathbf{E}=(E_1, E_2, E_3)$, and the strain ($\varepsilon_{jk}$) tensors, where i, j, k ∈ {1,2,3}, with 1, 2, and 3 corresponding to x, y, and z, respectively. (Further developments would be required to estimate the magnitude of the error associated with the inherent limitations to the presented linear approach[52]). PZ effect is described using the reduced piezoelectric tensors $d_{11} = \varepsilon_{11}/E_1$ and $d_{22} = \varepsilon_{22}/E_2$[16]. For the evaluation of PZ components, $d_{11}$ and $d_{22}$, we proceed through two main computational steps: (i) for a fixed $\mathbf{E}$-component we calculate in-plane strain (XY-plane according to representation in Fig.1) associated to each chemical bond in the selected reduced crystal (shadow region in Fig.1) (ii) for the extension of such reduced models to the extend crystal, we follow a procedure similar to the one already proposed for a similar hydrogen-bonded crystal[14]. Numerical results for PZ values (the slopes of the curves in the linear regression models) are reported with a standard error lower than 15 % of the values associated with the corresponding slopes, see Supp. Mat. 4 for further details.

Full version is available on: https://www.nature.com/articles/srep34616

AUTHOR INFORMATION

**Corresponding Author**

Correspondence and requests for materials should be addressed to: yasserblanco@sce.carleton.ca or clivia.sotomayor@icn2.cat.

**Author Contributions** Y.G. and Y.B.R-B. designed the research plan, developed the electromechanical models and carried out numerical calculations. Y.M-P. supervised the analysis of such aspects related with the definition and interpretation of the numerical descriptor. C.M. S-T. conceived the project and directed the research work. All authors discussed the results and contributed to the manuscript writing.

**Funding Sources**

This work was supported by the Spanish MINECO-project TAPHOR: MAT2012-31392 and the catalan AGAUR project 2014 SGR 1238.

**Notes**

Supplementary material is available on the online version of the manuscript.

**Acknowledgments.** Instructive comments from Jouni Ahopelto and Gustau Catalan are acknowledged. We are also grateful with Prof. Yoshiharu Nishiyama for his help on the interpretation of X-rays data. Y. García is indebted with Andrés Rodríguez for the thoughtful motivational discussions regarding HB intricacies. Computational facilities have been provided by "Centro de Supercomputación de Galicia" (CESGA) and ICN2. This work was supported by the Spanish MINECO-project TAPHOR: MAT2012-31392 and the catalan AGAUR project 2014 SGR 1238.

Full version is available on: https://www.nature.com/articles/srep34616

**Competing Financial Interests**

We have filed an internal Notification of Invention based on this work while considering patenting.

**Figure´s list**

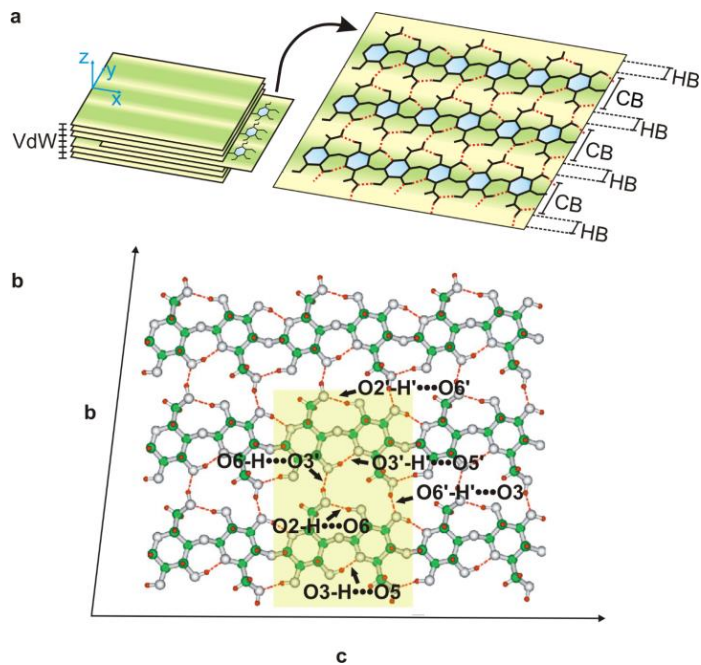

**Figure 1. Hydrogen bonds pattern in Iβ-NC. a**, Schematic representation of relevant chemical bonds and atomic arrangements in NC. Quasi-2D slabs in NC are stacked by VdW interactions. HB and covalent bonds are represented by red and black colours, respectively. **b**, Atomistic representation of the HB pattern in NC. In green, grey and red colours are represented carbon, oxygen and hydrogen atoms. The HB pattern is represented by red dashed lines. The model cluster comprises three chains linked by inter-chains HB. The NC chains are composed of six glucose monomers linked by glycosidic bonds whose torsion angles are modulated by two intra-chain HB. The region within the light-green rectangular region is the smallest cluster defined to study HB effects. Within the six HB contained in the cluster, there are 3 pairs of equivalent HB associated

to equivalent atoms of different chains: [O3-H···O5 ~ O3'-H'···O5'], [O2-H···O6 ~ O2'-H'···O6'] and [O6-H···O3' ~ O6'-H'···O3].

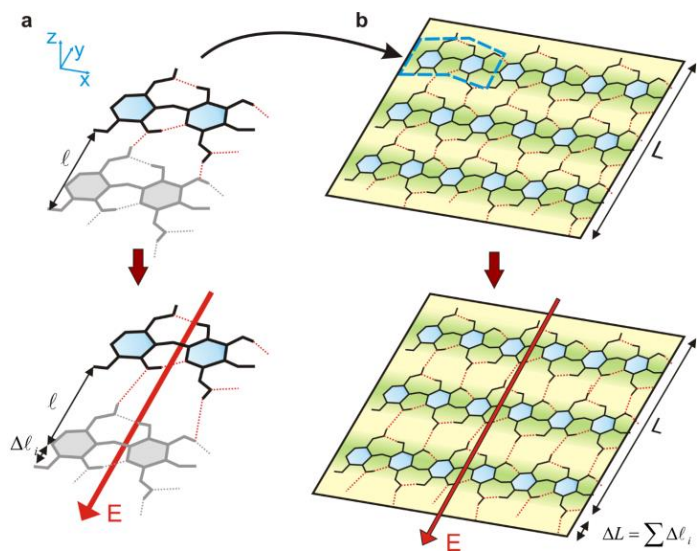

**Figure 2. Illustration of the atomic scale mechanism of piezoelectricity in Iβ-NC**. Right panels are reduced regions representing the larger NC slab. The left side depicts schemes for the smallest unit needed to explain the molecular mechanism of the piezoelectric response. **a,** The distance between equivalent atoms of each chain is represented by '$l$'. **b,** The model units are held along the b-axis direction by HB, whereas the covalent chains are oriented in the c-axis direction. The CB tightens the system constraining electromechanical responses in c-axis direction. 'L' denotes the length of the crystal in the b-axis. **c,** Effect of an **E**-field in the b-direction. The variation in 'L' due to an external field is represented by a finite $\Delta l_i$ value in this direction. **d,** Such atomic-scale effect is extended along the *bc* plane in the b-direction of the NC crystal. Then, in terms of collective effects we predict stretching and compression behaviour in this direction.

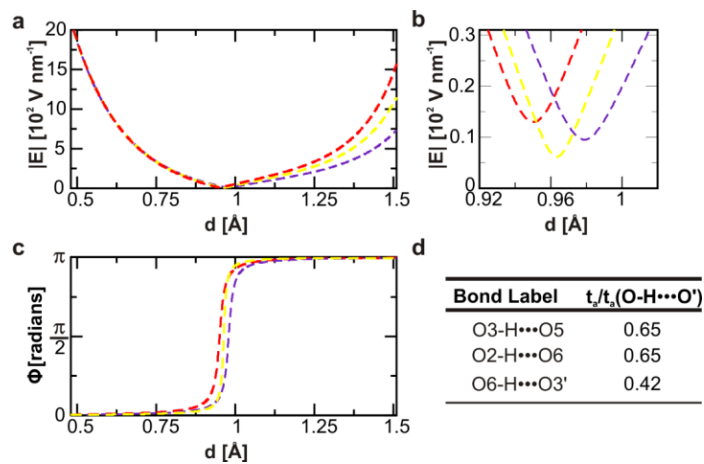

**Figure 3. Electrical characterization of representative HB.** Three representative HB are characterized. Distinctive HB in Iβ-NC are the intra-chain O3-H⋯O5 and O2-H⋯O6, as well as the inter-chain O6-H⋯O3'. Such three bonds are coloured in red, indigo and yellow, respectively. In **a**, and **b**, we represent the variations of the electric field modulus (|**E**|) along the interaction axis (O → H). The asymptotic behaviour coincides with atomic positions with positive charged centres. The existence of finite threshold values ($E_{threshold}$ ~ 10 V nm$^{-1}$) is essential for understanding the electrostatic inertia as well as the static response originated in the NC crystal. **c** and **d**, show the characteristic evolution of the angle between the **E** vector and the two atom axis, Φ= Φ(d). The deviations of such curves from the limit Heaviside function are indicative of the electrostatic instability of chemical bonds. The descriptors included in the table accounts for this information. The numerical values, normalized to the HB in the water dimer (O-H⋯O') are indicative of a lower electrical activity in NC than in the water dimer due to the implication of HB, see also Supp. Mat. 2.

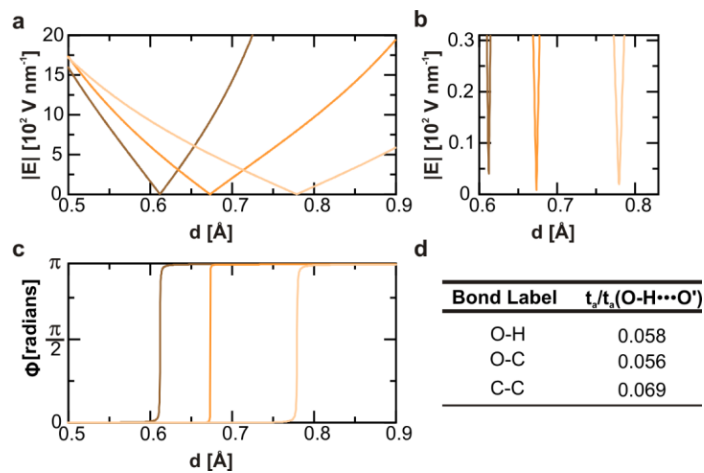

**Figure 4. Electrical characterization of representative CB.** Three representative CB are studied, O-H (represented in black colour), O-C (in brown), and C-C (in grey). Apart from such generalities already described for HB, see Fig. 3, here **E** gradient are stronger than in the case of HB leading to a reduced electric susceptibility of such bonds in the presence of neighbour HB. This information is also captured by the numerical index, $t_a$, obtained from the analysis of the evolution of the angle between the **E** vector and the interaction axis, $\Phi= \Phi(d)$, in **c**. The descriptor values are included in the inset table. The $t_a$ values normalized to the corresponding value of the HB in the water dimer provide a measurement of the significantly lower electric susceptibility of CB in NC backbones related to the HB of the water dimer taken as reference.

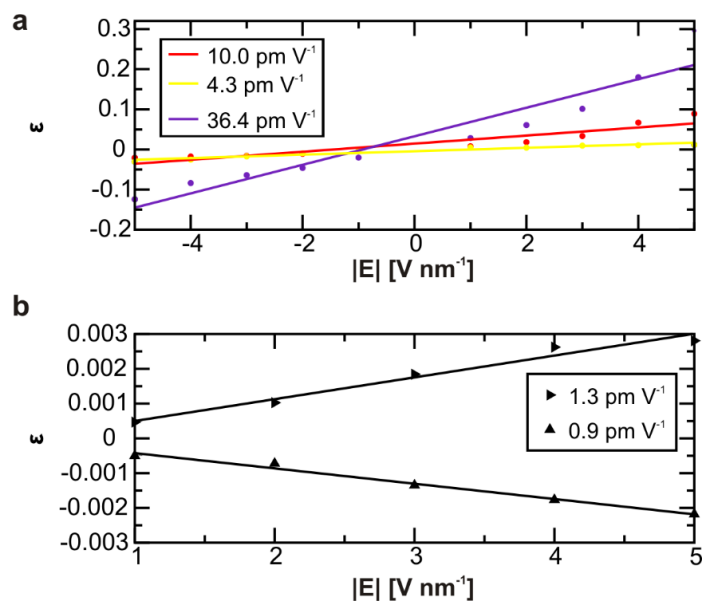

**Figure 5. Piezoelectric responses in 2D-NC estimated from atomic scale calculations. a,** Dependence between the HB strain ($\varepsilon$) and the electric field intensity (**E**) when both are aligned with an interaction axis with its origin in the oxygen atom, O → H. Results for the intra-chain HB O3-H⋯O5 and O2-H⋯O6, are represented by red and indigo colours, respectively, and for the inter-chain HB O6-H⋯O3' in yellow. Numerical outcomes are indicated by dots. The slopes of the linear regression curves, shown in the right panel, are indicative of the bond oriented piezoelectric coefficients for the three representative HB, see Supp. Mat. 4. **b** Estimated 2D-NC total bond localized contributions to strain as **E** rotates in the XY plane. By right-oriented arrows we indicate total PZ effect when **E** is oriented along the y-axis. By upwards-oriented arrows we show the results when **E** rotates 90º. The linear interpolation curves leads to orthogonal PZ coefficient values that represent the sum of the single bond localized contributions in a reduced representation of the crystal, see shadow region in Fig. 1b. The extrapolation of such coefficients to the extend crystal leads to the estimated PZ coefficients of 2D-NC. The orthotropic behaviour in NC is evidenced by the 30% relative variation for the two PZ tensor component values for the extended crystal, $d_{22}$ and $d_{11}$ respectively, reported in the right panel, further details in Methods section.